\documentclass{iau} 
\usepackage{graphicx}

\title[Automated Detection Methods for Solar Activities] 
{Automated Detection Methods for Solar Activities and an Application for Statistic Analysis of Solar Filament}

\author[Q. Hao \& P. F. Chen \& C. Fang]   
{Q. Hao$^{1,2}$
\and P. F. Chen$^{1,2}$
 \and C. Fang$^{1,2}$
 }

\affiliation{$^1$School of Astronomy and Space Science, Nanjing University, Nanjing 210023, China \\ email: {\tt haoqi@nju.edu.cn} \\[\affilskip]
$^2$Key Laboratory of Modern Astronomy and Astrophysics (Nanjing University), Ministry of Education, China}

\pubyear{2018}
\volume{340}  
\jname{Long-Term Datasets for the Understanding of Solar and Stellar Magnetic Cycle}
\editors{D. Banerjee, J. Jiang, K. Kusano \& S. Solanki eds.}

\usepackage{graphicx}
\usepackage{pdfpages}
\usepackage{epsfig,natbib,color}
\usepackage{lscape}
\usepackage{graphicx}
\usepackage{epsfig}
\usepackage{natbib}
\usepackage{gensymb}
\usepackage{multirow}

\newcommand\apj{{Astrophys. J.}}
\newcommand\apjs{{Astrophys. J. Suppl. Ser.}}
\newcommand\aap{{Astron. Astrophys.}}
\newcommand\mnras{{Mon. Not. Roy. Astron. Soc.}}
\newcommand\solphys{{Sol. Phys.}}

\begin{document}

\maketitle

\begin{abstract}
With the rapid development of telescopes, both temporal cadence and the spatial resolution of observations are increasing. This in turn generates vast amount of data, which can be efficiently searched only with automated detections in order to derive the features of interest in the observations. A number of automated detection methods and algorithms have been developed for solar activities, based on the image processing and machine learning techniques. In this paper, after briefly reviewing some automated detection methods, we describe our efficient and versatile automated detection method for solar filaments. It is able not only to recognize filaments, determine the features such as the position, area, spine, and other relevant parameters, but also to trace the daily evolution of the filaments. It is applied to process the full disk H$\alpha$ data observed in nearly three solar cycles, and some statistic results are presented.
\end{abstract}

\firstsection 
\section{Introduction}

As the result of solar activities such as flares, filament eruptions, and coronal mass ejections (CMEs) \citep{Chen2011}, severe space weather in the near Earth space has societal and economic effects on human systems. In order to monitor and analyze solar activities, many ground-based and space-born telescopes have been built, with both the time cadence and the spatial resolution becoming higher and higher. As a consequence, we have to deal with a vast amount of data, and automated detection is an efficient way to derive the features of interest in the observations. In the past decades, automated detection attracted a lot of attention in the solar physics community. Researchers used different image processing and machine learning applications to obtain reliable results, which significantly expedited their research.

In this paper, we present a brief overview of automated methods developed for the detection of sunspots, solar flares, CMEs, and filaments, and then summarize the results obtained with the filament detection method developed in our group. It should be noted that the automated detection has also been used in identifying other solar phenomena, such as active regions \citep{Zhang2010,Caballero2014}, coronal EIT waves \citep{Podladchikova2005,Long2014}, oscillations \citep{Sych2010}, coronal holes \citep{Scholl2008,Krista2009,Kirk2009}, coronal loops \citep{McAteer2010}, and small-scale structures like magnetic bright points \citep{Crockett2009,Javaherian2014}, granules \citep{Feng2012,Yang2015}, and chromospheric fibrils \citep{Schad2017}, which are beyond the scope of this paper.

\section{Overview of automated detection methods for solar activities
}

{\underline{\it Sunspots}}.
Sunspots are the most noticeable phenomenon on the solar surface. They are generally observed in white-light as dark areas in contrast to the brighter quiet Sun. Consequently, early attempts for sunspot detections were focused on selecting a suitable threshold to distinguish the umbra from the penumbra and from the penumbra from the quiet Sun. However, the sunspot areas are often underestimated because of some bright dots inside the sunspots. In order to solve this problem, some other methods  based on the morphological operations were developed. For example, \citet{Zharkov2005a} used the edge-detection method and a local threshold to find a sunspot candidate, then employed a median filter  to remedy the possible over-segmentation of a sunspot because of the intensity inhomogeneity, and \citet{Zhao2016} used Otsu algorithm to find an adaptive threshold for the sunspot segmentation. Note that a top-hat operation is often conducted before segmentation in order to enhance the contrast between sunspots and the background \citep{Curto2008, Watson2009, Pucha2016}.

Some other techniques have also been implemented. \citet{Turmon2002} adopted a statistical Bayesian technique for the sunspot detection. \citet{Colak2008} applied both intensity threshold and the region-growing technique for the initial detection of sunspot regions and then adopted the neural network technique for their classification. \citet{Fonte2009} determined the sunspot umbra and penumbra boundaries based on the fuzzy set theory. \citet{Djafer2012} applied compact wavelet transform to automatically identify sunspots. \citet{Goel2014} adopted a method called level-set image segmentation which takes advantage of both the image gradient and the region-based statistics to detect sunspots.

{\underline{\it Solar flares}}. 
Solar flares are one of the most energetic phenomena that take place in the solar atmosphere \citep{Shibata2011}. Their detection is a key part in space weather monitoring. The longest continual data for solar flares are the GOES X-ray flux. \citet{Aschwanden2012} analyzed the soft X-ray light curves over 37 years and developed an automated flare detection algorithm. With the imaging observations available, we can further obtain the locations and morphologies of the flares. \citet{Veronig2000b} utilized the region-based and edge-based segmentation methods to track the flare ribbon separation. An image-processing method based on active contours was proposed to track UV and EUV flare ribbons \citep{Gill2010}.  With the region-growing, morphology and motion tracking techniques, the evolution of flare ribbons can be measured  in an entirely automatic way \citep{Qu2004,Maurya2010}. \citet{Kirk2013} proposed an algorithm to detect and track flare ribbons and sequential chromospheric brightenings in H$\alpha$. \citet{Borda2001} described a method for the automatic detection of solar flares using the multi-layer perceptron (MLP) and used a supervised learning technique that required a large number of iterations. \citet{Qu2003} compared MLP, radial basis function and support vector machine (SVM) for solar flare detection on solar H$\alpha$ images. The experimental results show that by using SVM they obtained the best classification rate. \citet{Mravcova2017} developed a code based on the breadth-first search algorithm to automatically detect the kernels of white-light flares observed in the \textit{SDO}/HMI intensity maps. The automatic detection of solar flares is also beneficial for the satellite observations to be shifted to the high-cadence mode or from the full-disk to partial-disk mode. Moreover, the detection of precursors for solar flares would be crucial in space weather forecast \citep{Wang2003}.

{\underline{\it CMEs}}. 
CMEs have a close relationship with many other solar eruptive events and are the major driving source of the hazardous space weather around the Earth \citep{Chen2011}. The fastest CMEs can reach the Earth within a day, and the slower ones take up to 4 or 5 days to reach the Earth. The automated detection methods can quickly estimate the speed of a CME and its arrival time at 1 AU, which allows us to take necessary actions in time. By applying the modified Hough transform to the running difference images of the SOHO/LASCO coronagraph observations, \citet{Berghmans2002} proposed an automatic detection method for CMEs, which is called ``Computer Aided CME Tracking (CACTus)". \citet{Robbrecht2004} improved CACTus, thereby the success rate was improved from $\sim75\%$ to 94\%. They also identified some CME events that have not been listed on the existing human-vision-based CME catalog. Similarly, \citet{Tappin2012} developed a method called ``Automatic Coronal Mass Ejection Detection tool (AICMED)", where they also applied the Hough transform to the time-elongation J-maps of the Solar Mass Ejection Imager (SMEI) data. \citet{Pant2016} used a modified version of the CACTus and applied it to automatically detect interplanetary CMEs observed by \textit{STEREO}/HI-1.

Later, \citet{Qu2006} tried to detect CMEs by applying a variable intensity threshold to the running-difference images observed by SOHO/LASCO C2 and C3 coronagraphs, then used a morphological operation to recognize the CME features in order to classify CMEs into different categories using SVM technique. \citet{Olmedo2008} developed a routine called Solar Eruptive Event Detection System (SEEDS), which detects CMEs in the polar transformed running difference images constructed from the SOHO/LASCO C2 data. The detection process is based on a threshold-segmentation technique using a region-growing algorithm and morphological operations. \citet{Boursier2009} developed an automatic method for detecting CMEs from synoptic maps, called Automatic Recognition of Transient Events and Marseille Inventory from Synoptic maps (ARTEMIS), which involves adaptive filtering, segmentation and merging with experiential knowledge. \citet{Morgan2012} and \citet{Byrne2012} developed the Coronal Image Processing (CORIMP) method. Rather than using a running difference technique to remove the unchanged background coronal structures, they developed a new deconvolution technique that separates coronagraph images into static and dynamic components. They found that such a technique can recognize faint CMEs. Recently, \citet{Hutton2017} developed a method which can detect the CME structures in three dimensions based on the observations from two vantage points, such as the SOHO/LASCO C2 and \textit{STEREO} COR2 data.

{\underline{\it Solar filaments}}. 
Solar filaments are prominences projected against the solar disk and particularly visible in H$\alpha$ observations, where they often appear as elongated dark spines with several barbs \citep{Tandberg1995}. Filament eruptions are often associated with flares and CMEs and therefore are a key ingredient space weather \citep{Chen2011}. Like sunspots, filaments are also dark features on the solar disk, so the segmentation methods for sunspots are sometimes applicable for filaments. \citet{Gao2002} first used a global threshold and region-growing techniques to detect filaments. \citet{Shih2003} adopted local thresholds which were chosen by median values of the image intensity to extract filaments. However, this kind of threshold selection cannot guarantee robust results since the bright features on images can significantly affect the value of the thresholds associated with the median value. Therefore, in order to overcome this problem, some authors developed adaptive threshold methods \citep{Qu2005, Yuan2011, Hao2015}. Particularly, \citet{Qu2005} applied the SVM technique to distinguish filaments from sunspots.

A filament may be split into several fragments during its evolution. To figure out whether they belong to one filament is crucial for the study of filament evolution. Initially, many authors adopted morphological operations to connect the fragments \citep{Shih2003, Fuller2005}. However, the results depend on the selection of the structure element. Later, a distance criterion was used to identify the fragments belonging to a single filament \citep{Gao2002, Joshi2010, Hao2013}. However, in some cases several filaments in an active region are so close to each other that they satisfy the distance criterion and would be recognized as one filament. To avoid this error, \citet{Bernasconi2005} and \cite{Hao2015}  added the slopes of the fragments. If the fragments satisfy the distance criterion and also have a similar slope, they would be treated as one filament. \citet{Qu2005} used an adaptive edge linking method to connect filament fragments, which is based on a similar concept. \citet{Bonnin2013} developed an algorithm by comparing the shapes of all the possible spines.

With relatively high-resolution images in H$\alpha$, we can see that a filament is composed by a spine and many barbs. In order to derive the filament spine, many authors employed iterations of the morphological thinning and spur removal operations \citep{Fuller2005, Qu2005, Hao2013}. However, after iterated spur removal operations the spines often become shorter than the original ones. \citet{Bernasconi2005} used the Euclidean distance method to find the end points of the filament skeleton in order to derive the real filament spine. \citet{Yuan2011} and \citet{Hao2015,Hao2016} used the algorithm based on the graph theory by calculating the number of pixels on the path from one end point to the other. The longest path between a pair of end points is kept as the main skeleton.

As for prominences, they always appear above the solar limb, therefore, it is relatively easier to detect them since the background is the faint corona. The difficulty is how to distinguish the prominences from active regions and coronal loops since they all are bright features. \citet{Foullon2006} implemented a histogram segmentation to set an intensity threshold for an individual image and then employed multi-wavelength observations to exclude active regions. \citet{Wang2010} first selected potential prominences by a certain threshold and then performed a linear discriminant analysis based on the shape of the selected regions to exclude the non-prominence features. \citet{Labrosse2010} used moments to find radial profile characteristics and the SVM method to determine which feature is a prominence. Then they reconstructed the shape of the prominence by morphological opening procedures. \citet{Loboda2015} identified a feature to be a prominence based on its characteristic in the He \small{I} 304 \AA.

\begin{figure}[b]
\begin{center}
 \includegraphics[width=0.7\textwidth,clip=]{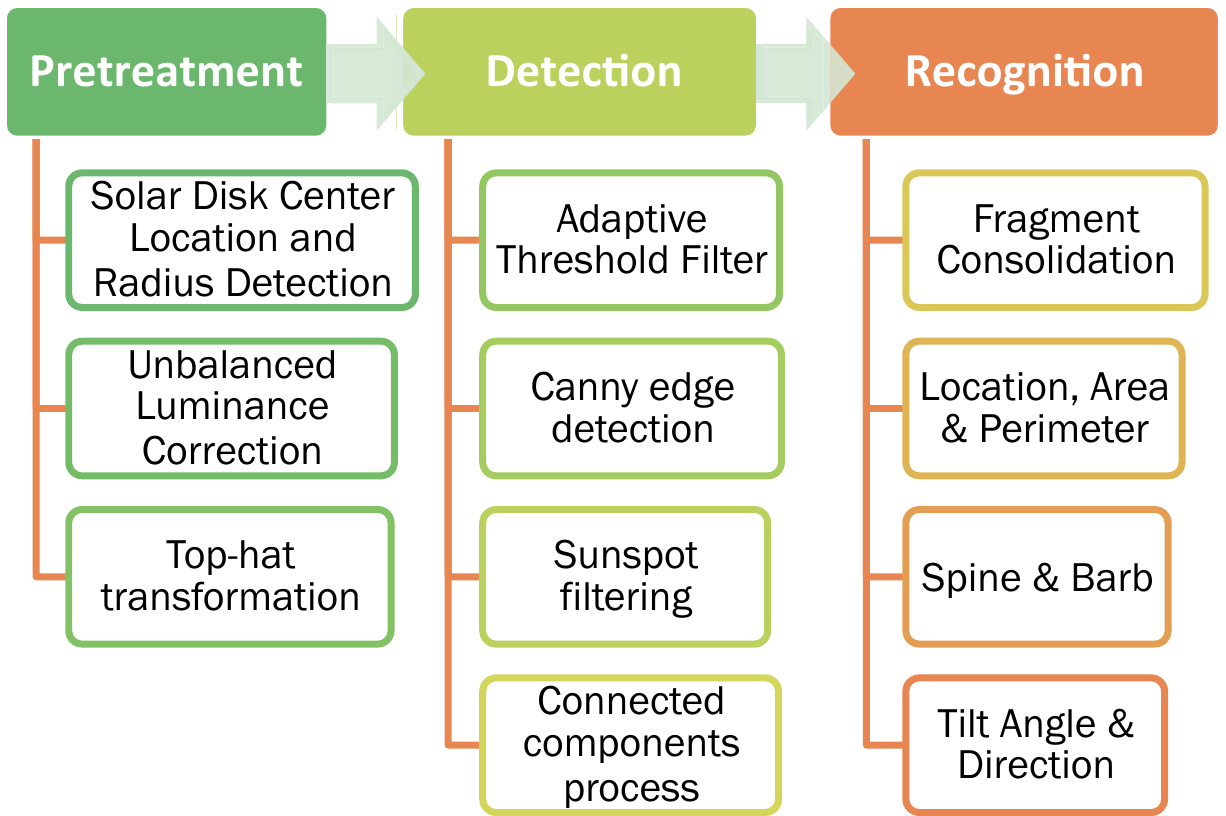} 
 \caption{The flowchart of our filament automated detection method \citep{Hao2015}.}
   \label{fig1}
\end{center}
\end{figure}

\begin{figure}[b]
\begin{center}
 \includegraphics[width=1.0\textwidth,clip=]{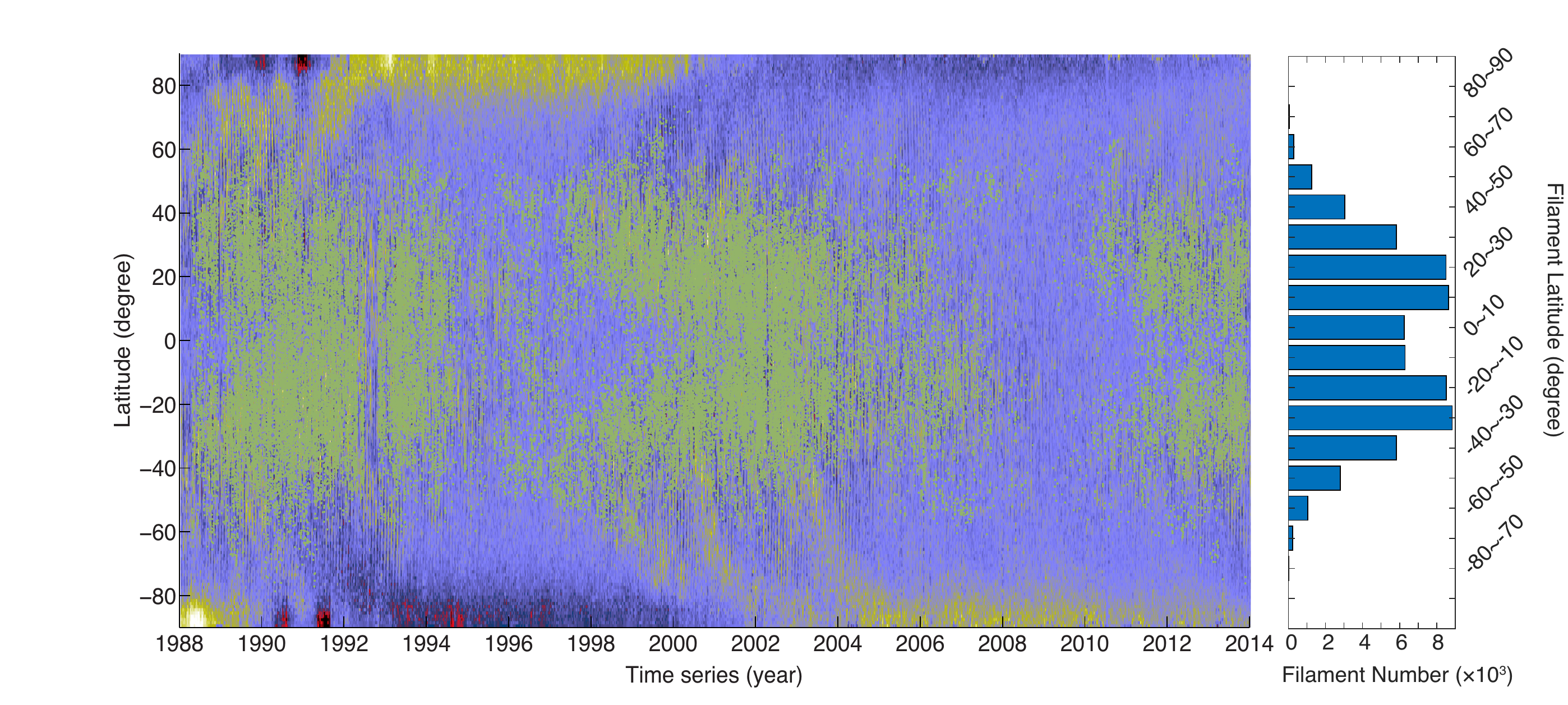} 
 \caption{Left: Butterfly diagram of filaments from 1988 to 2013. Each green dot represents a single observation. The background is the butterfly diagram of magnetic fields during the same time periods for comparison. Right: Distributions of the Detected filament numbers in different latitude bands.}
   \label{fig2}
\end{center}
\end{figure}

\section{An application for solar filaments statistic analysis}

We developed and improved an automated detection and tracing method for solar filaments \citep{Hao2013, Hao2015}. The method mainly has three parts: pretreatment, detection of filaments, and feature recognition. The corresponding flowchart is shown in Figure~\ref{fig1}. In the pretreatment step, we preprocess the raw data, such as the limb-darkening correction and image enhancement. Then, we adopt the Cannny edge detection technique to identify individual filaments. In the meantime, sunspots are excluded. After this step, each filament is marked with a unique identity number. Finally, we employ the morphological operations to recognize the filament features, such as their positions, directions, spines, areas, perimeters, and so on. We collected H$\alpha$ data from various observatories to test our method and it was demonstrated to be very efficient and universal. 

We applied our method to the data mainly observed by the Big Bear Solar Observatory during the period from 1988 to 2013 \citep{Hao2015}. The temporal evolution of the latitudinal distribution of these filaments, known as the ``Butterfly diagram", is plotted as the scatter plot in Figure~\ref{fig2}. Each green dot represents a single observation. For comparison, the background is the butterfly diagram of the magnetic field during the same time periods. From the diagram we can see the distribution and the migration of the filaments. The right panel of Figure~\ref{fig2} shows the latitudinal distribution of the detected filament numbers within the three solar cycles. The latitudinal distribution of the filament numbers is bimodal. The peak values are within the latitude band [$10^{\circ}$, $30^{\circ}$] in both hemispheres. We also analyzed the filament area, spine length, tilt angle and barb numbers.

We calculated the monthly mean latitude of the filaments in the northern and southern hemispheres in order to examine filament migrations. A cubic polynomial fitting was employed to derive the drift velocity. The fitting results show that the monthly mean latitude of the filaments has three drift trends: from beginning to the solar maximum of a solar cycle, the drift velocity is very fast; after the solar maximum it becomes relatively slow; near the end of the solar cycle, the drift velocity becomes divergent.

We also calculated the north-south (N-S) asymmetry of the filament features. The N-S asymmetry indices of the filament numbers, filament numbers with various areas, spine lengths, and the cumulative areas and spine lengths indicate that the southern hemisphere is the dominant one in solar cycle 22 and the northern hemisphere is the dominant one in solar cycle 23. Though the difference between the two hemispheres is not significant. The N-S asymmetry indices show that the northern hemisphere dominates in the rising phase of solar cycle 24.

\section{Prospects}
Solar activities, such as flares, filament eruptions, and CMEs, can lead hazardous space weather near the Earth. Automated detection methods not only expedite the research regarding the mechanisms of these phenomena, but also can make alert and forecast of hazardous the space weather. Despite the detection success accuracy is not $100\%$ at the moment, with the development of data processing techniques, especially the artificial intelligence methods, automated detection of solar activities can lead solar physics to a new stage.\\

{\underline{\it Acknowledgments.}} The research was supported by NSFC (grants 11533005, 11703012, and 11733003), NKBRSF (grant 2014CB744203), Jiangsu NSF (grant BK20170629), and Jiangsu 333 project. 


\end{document}